\documentclass[acmlarge]{acmart}
\AtBeginDocument{%
  \providecommand\BibTeX{{%
    \normalfont B\kern-0.5em{\scshape i\kern-0.25em b}\kern-0.8em\TeX}}}

\setcopyright{acmcopyright}
\copyrightyear{2018}
\acmYear{2018}
\acmDOI{10.1145/1122445.1122456}

\begin{document}

\title{SonarWatch: Field sensing technique for smartwatches based on ultrasound and motion}

\author{Yingtian Shi}
\affiliation{%
  \institution{Tsinghua University}
  \country{China}
}

\author{Chun Yu}
\authornote{Corresponding author.}
\affiliation{%
  \institution{Tsinghua University}
  \country{China}
}

\author{Xuyang Lu}
\affiliation{%
  \institution{Tsinghua University}
  \country{China}
}
\author{Xing-Dong Yang}
\affiliation{%
  \institution{Simon Fraser University,}
  \country{Canada}
}

\author{Yuntao Wang}
\affiliation{%
  \institution{Tsinghua University}
  \country{China}
}
\author{Yuanchun Shi}
\affiliation{%
  \institution{Tsinghua University}
  \country{China}
}


\def \projectName {{SonarWatch}}

\begin{abstract}

A smartwatch, as an electronic device worn continuously on the wrist, has the potential to perceive rich interactive gestures and natural behaviors of the user. Unfortunately, the current interaction functionality of smartwatches is primarily limited by the sensing ability of a small touch screen. This paper proposes the \projectName{}, a novel sensing technique that uses the acoustic field generated by the transceiver on the opposite sides of the watch to detect the presence of nearby objects and has the potential to infer their shapes. This enables a range of same-side hand interactions, opposite-side hand interactions, and natural behavior perception. We designed an algorithm combining IMU and acoustic fields to identify these actions and optimize power consumption. We tested the performance of \projectName{} in different distances and noise environments, achieving an overall accuracy of 93.7\% based on the data from 12 participants, with the accuracy of same-side hand interaction at 97.6\%, opposite-side hand interaction at 95.76\%, and body or object interaction at 99.1\%. Its power consumption is close to that of physiological sensors. \projectName{} can achieve the above capabilities by utilizing the existing built-in sensors, such as in the Apple Watch, making it a technology with solid practical value.

\end{abstract}




\begin{teaserfigure}
  \includegraphics[width=\textwidth]{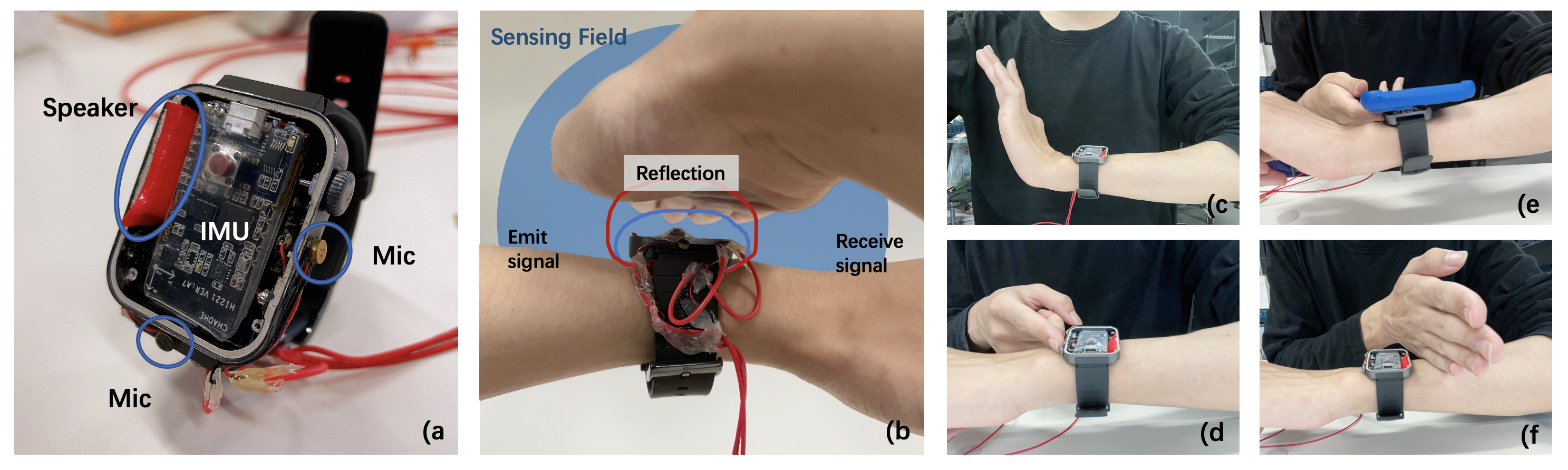}
  \caption{\projectName{} concept: a) Sensors used by \projectName{} and their distribution b) Sensing principle: user gestures affect the sound propagation path to provide features for recognition. The right figure is an example of the interaction space: c) Same-side gesture: Wrist Up; d) Opposite-side touch gesture: Click Mic; e) Special gesture: Phone close to Watch; f) Opposite-side touchless gesture: Block Left; }
  \label{fig:teaser}
\end{teaserfigure}

\maketitle
\section{Introduction}
Smartwatches have gained significant traction among consumers in recent years due to their convenience and portability. The smartwatch is worn on the user's wrist. It follows the movement of the user's arm, which has great potential to obtain information about the user's gestures and postures. However, watch interaction methods are primarily limited to touchscreens. When the screen is too small, users may find it challenging to perform accurate taps, leading to errors and frustration. Moreover, the touch input may not be practical when users' hands are restricted, such as during driving or exercise. The rich information provided by the user's postures and gestures cannot be perceived, significantly affecting the watch's interaction ability and the user's experience.

The position and motion of the user's arm and hand constitute an interaction space around the watch. Past research only explored subsets of this space to enhance input modalities on the watch. They introduced touch interactions beyond touchscreens\cite{sidesight, skinbutton}, same-side hand gestures like pinch \cite{pinchwatch, serendipity, zhang2018fingerping, zhang2017fingersound}, and opposite-side hand gestures such as clap \cite{gestear} and waving\cite{ruan2016audiogest, gupta2012soundwave}. At the same time, implementing these interaction methods requires additional sensors such as microphone arrays, optical sensors, and magnetic sensors, resulting in bulky and inconvenient smartwatches that are challenging to wear and use.

Based on interaction space around smartwatches, we present \projectName{}, a novel sensing technique for enhancing the interaction capabilities of smartwatches. \projectName{} uses the watch's built-in sensors to create a sound field, which can sense nearby objects and their shapes. This field following the wrist extends the sensing range of the smartwatch to the surrounding space. By combining the data from the acoustic and motion sensors, \projectName{} significantly improves the sensing space of the smartwatch.

We re-outline the interaction space of the smartwatch into three subspaces: Same-side Hand Gesture, Opposite-side Hand Gesture, and Body or Object Interaction. We propose a number of novel interaction techniques to demonstrate the power of \projectName{}. With the data from 12 participants, our study shows that \projectName{} achieved recognition accuracy of 93.7\% across 12 classifications. The accuracy of \projectName{} in recognizing the opposite-side gesture and body or object interaction is 95.76\% and 99.1\%. Notably, when tilting the wrist upwards, the accuracy of the five types of same-side hand gestures was as high as 97.6\%. 

We selected different gesture sets and sensor combinations to test the sensing range and sensitivity of \projectName{}. In particular, \projectName{} uses the watch's built-in sensors without relying on additional sensing hardware. To thoroughly test the usability of \projectName{}, we designed the algorithm to save energy and tested \projectName{}'s anti-noise ability, false touch, and energy consumption. The test results prove that \projectName{} has high practicability and value of direct deployment.

In summary, there are three main contributions in this paper:
\begin{enumerate}
    \item We propose \projectName{}, a smartwatch interaction technique based on the sound field. \projectName{} has rich perception capabilities to meet various interaction needs of users and does not rely on high-cost sensors and computing units.
    \item 
    We re-outlined the interaction design space around smartwatches and suggested a series of interaction gestures that offer additional design insights for field-based interaction technologies.
    \item We conducted a targeted study focusing on the practicality of a specific implementation of acoustic field sensing techniques.
     We tested the sensing range and sensitivity of \projectName{} and analyzed its energy consumption and false positives.
\end{enumerate}

\section{Related Work}

\subsection{Smart Watch Interaction Modalities Enhancement}
Touch interaction on mobile devices (e.g., mobile phones, watches) has been widely studied\cite{hightouch, SelectionMultiTouch, accuracyTouch, fingerangle}, 
and the ultra-small screen of smartwatches brings more apparent problems of "fat fingers" and occlusion. 
In the two-dimensional Fitts experiment, the click error rate is as high as 66\% when the target width is 2.4mm\cite{bi2013ffitts}. 

To surpass these challenges and provide more effective interaction solutions, researchers have explored smartwatch interactions beyond touchscreen.
\citet{arefin2016exploring} introduced a repertoire of 18 gestures that extend beyond the scope of traditional touchscreen inputs. Loclair C\cite{pinchwatch} first proposed using a chest camera to detect the user's pinch. Researches \cite{WristFlex, zhang2018fingerping, zhang2017fingersound} also explored the pinch using different sensor combinations.
Sidesight \cite{ sidesight}, and Skinbutton \cite{skinbutton} utilize infrared sensors on the watch side to detect user clicks on the skin. Researches \cite{touchEdge, expandWatch} explore touch interaction with the watch edge using a capacitive sensor or microphone. Both WristWhirl\cite{wristwhirl} and WrisText\cite{wristext} use a complex combination of proximity light sensors, piezoelectric sensors, and IMUs to identify the user's wrist movements. However, the interaction possibilities explored by these studies were often constrained, leaving significant portions of the watch interaction space untapped.

While previous studies exhibited a rich variety of interaction modalities, their efficacy was often impeded by the limited sensing range of their sensors. By conducting a comparative analysis, it becomes evident that field sensing technologies based on magnetic, light, and sound fields offer superior sensing capabilities.
BeamBand\cite{iravantchi2019beamband} migrates the sonar sensing technique to the watch. 
The researchers added an array of ultrasonic sonars to the wristband, which can recognize the different gestures of the user's palm. 
Harrison C\cite{abracadabra} deploys magnetic sensors on the user's fingertips and the watch to expand the interactive area of the screen, which helps users perform mid-air gestures in a large space.
The gesture watch \cite{gestureWatch} uses infrared sensors to detect mid-air gestures such as Waves or Cover of the user's opposite-side hand to the watch.

However, these studies often incorporated additional sensors to achieve interaction diversity. 
While these endeavors augmented the watch's interaction potential, the additional sensors make these solutions difficult to apply in practice and make the watch inconvenient to wear. Building an effective sensing field that does not rely on large, hard-to-deploy sensors is meaningful and can complete the sensation of more interactions.
Therefore, devising an effective sensing field that doesn't rely on large and unwieldy sensors presents a meaningful avenue for enriching the feasible interactions without compromising wearability and convenience.

\subsection{Sensing Field for Interaction Improvement}
The features of various fields like magnetic, sound, and electric fields change in response to the characteristics and actions of objects within them, enabling the identification of interactive actions. Sensing methods based on the field have been widely used in many interactive technologies and practical applications due to their broad coverage and recognition capabilities.

For instance, in electric field research,
\citet{smith1998electric} uses electric fields to enhance the interaction of graphical interfaces. EarFieldSensing\cite{EarFieldSensing} utilized the electric field in the ear  for user expression detection. Tomo\cite{Tomo} uses electric field-based Electrical Impedance Tomography (EIT) for gesture recognition based on the user's forearm characteristics.
However, implementing electric fields on small mobile devices is challenging due to hardware complexity. In contrast, acoustic and magnetic fields are better suited for mobile scenarios.

Interferi \cite{Interferi} recognizes gestures and facial expressions using ultrasound-based fields in the bracelet and mask. EchoFlex \cite{EchoFlex} uses the principle of ultrasound imaging to complete the recognition of 10 discrete gestures. Touch\'{e}\cite{Touche} explores how sound fields behave in interactions with objects and liquids. Different behaviors of users touching objects can affect sound propagation, and the researchers captured these changes to identify user interactions with different objects. Some studies have also demonstrated using magnetic \cite{abracadabra, synwatch} and light fields\cite{skinbutton, gestureWatch}.
Even though these methods can be used in mobile scenarios, applying additional sensors will change the comfort of the original device and public acceptability.

When considering the interaction of small devices like smartwatches, field sensing schemes based on additional sensors are improper to be implemented. 
Compared to other methods, the ultrasonic sound field is the most impractical sensing field that can be constructed based on existing sensors to reduce the cost of additional sensors. We want to explore how the watch's ultrasonic sensing field can improve watch-related interactive input.

\subsection{Acoustic Sensing Technique on Mobile Device}
Acoustic sensing techniques have a wide range of applications in mobile devices\cite{lu2009soundsense}, interactive objects\cite{sato2012touche, ono2013touch}, and IoT scenarios. The sound signal sensing principle is divided into contact conduction and air propagation. 

Contact conduction solution relies on additional sensors that are similar to high-frequency IMU. FingerPing\cite{zhang2018fingerping} and FingerSound\cite{zhang2017fingersound} use a contact microphone mounted on the watch and the rings to detect the user’s pinch. TapSkin\cite{zhang2016tapskin} uses the contact microphone and IMU on the watch to recognize the user's click on the skin near the watch. Acustico\cite{gong2020acustico} uses the microphone array on the wristband to capture the location of users' finger taps when their wrist is on the table. However, the disadvantage of contact conduction technology is that it can only detect contact signals, such as hand movements and vibrations, while wearing a watch but cannot identify non-contact information.

In contrast, air propagation solutions have a wider range. One typical sensing method is based on the principle of ultrasonic sonar, which needs a speaker to emit origin signals. Both AudioGest\cite{ruan2016audiogest} and SoundWave\cite{gupta2012soundwave} try to deploy a combination of speaker and microphone on terminal devices. When the users wave their arms or make gestures in the sound field, the microphone will receive different audio signals to complete the recognition; FingerIO\cite{nandakumar2016fingerio} uses this sensing method to achieve accurate tracking of finger click position. SornarID\cite{kim2022sonarid} also verified the feasibility of this solution on smartwatches.

When considering sensing solutions on smartwatches, the sound field offers significant practicality and an expanded range of sensing capabilities. 
However, in past studies, researchers have focused on changes within the sound field and ignored information about the field itself. Smartwatch, as a mobile device that follows the movement of the user's arm, its inherent movements related to gestures furnish an abundance of contextual information\cite{serendipity,wristext}. 
So we chose the combination of ultrasonic field and IMU as the final sensors to build the sensing method. We deeply explored the sensing capability of this sensor combination. We plan to rely on the original sensor of the watch to explore a more capable field-sensing method without bothering the user's daily experience.
\section{Hardware and Signal Analysis}
\label{hard}
This section will introduce the hardware system built for \projectName{}. We tested data from a pilot study to find potential features for gesture recognition. The principles and features will be described in detail below.

\subsection{Hardware Design}
\begin{figure}[htbp]
\centering
\includegraphics[width=10cm]{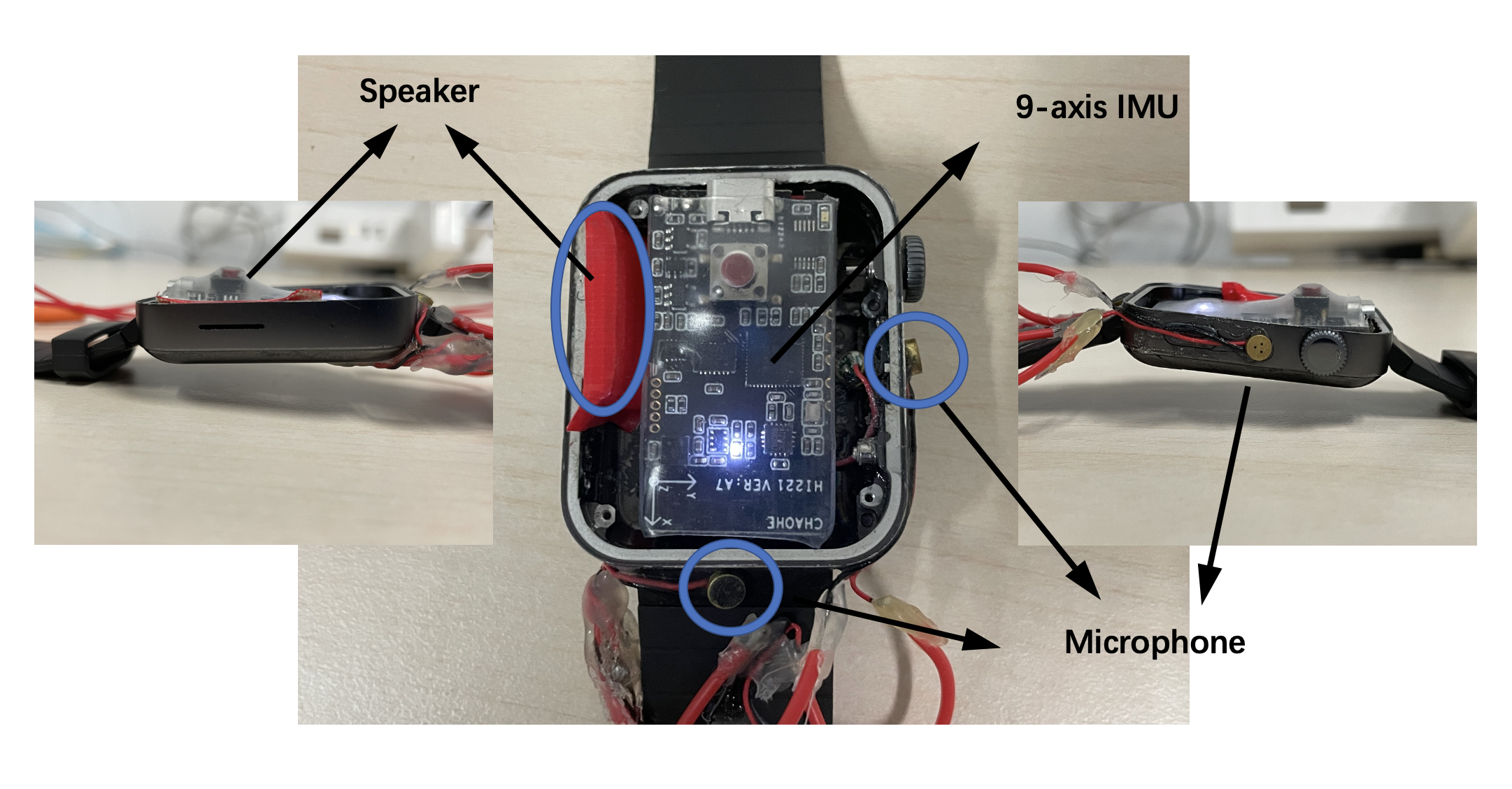}
\caption{The sensor distribution of the \projectName{}, consistent with the Mi watch, has a speaker on the left, a microphone on the right, and a nine-axis IMU inside. A microphone was added below the screen to explore more sensing possibilities}
\label{fig_3_h_1}
\end{figure}
We chose the microphone, speaker, and IMU as the sensors of the \projectName{} hardware system. The primary solution relies on the audio-sensing method, and the IMU provides movement-related recognition information. Considering that when using an existing smartwatch in the market, the microphone cannot obtain the original audio signal due to the AutoGain technique in the operating system. We selected sensors consistent with the Mi Watch and built a hardware system by referring to the sensor distribution of most smartwatches in the market (Fig. \ref{fig_3_h_1}). Based on the Mi Watch, we kept its shell and only kept the single speaker and microphone on the left and right sides of the shell. The position of the sensor is consistent with it on the Mi Watch, Apple Watch, and Samsung Galaxy Watch. Because of the design, there are different sensor distributions of watches on the market. For example, the microphone of the Huawei watch is located at the four o'clock position of the dial, and the LilyGo watch's microphone is at the back of the watch. Different placements and distribution of microphones can affect the system's sensing ability, so we added a microphone to the watch to explore more possible sensing opportunities of sound field sensing. Taking into account the watch's symmetry, placing the new microphone too close to the existing one would get similar signals. Thus, we positioned the new microphone below the screen. The microphone and speaker we used are the same as in the original Mi Watch. The released Apple Watch Ultra uses a three-mic array and dual speakers, demonstrating the possibility of using multiple microphones on smartwatches. We hope this research can provide a reference for the sensor distribution design of smartwatches. We will analyze the capabilities of the original sensor solution in the following section and explore the improvement of system capabilities by additional sensors. We embedded a 9-axis IMU (HI221) in the watch to capture the user's motion information, which is also consistent with the capabilities of the IMU on the Mi watch.
%

To enhance the anti-interference ability and feature richness of \projectName{}, we chosed up/down chirp signal (16.5kHz to 20kHz, sample rate 48kHz), which consists of multiple chunks. Each chunk contains 4096 sampling points (11.72 chunks per second, not the final sample rate). In the first half of each chunk, the signal frequency increases linearly from 16.5kHz to 20kHz, and in the second half, the signal frequency decreases linearly from 20kHz to 16.5kHz. The signal has good anti-interference performance against noise due to its high frequency. Its short period renders it adept at accommodating transient variations induced by gestures. Moreover, the signal's spectrum covers a wide range, affording it enough spectral features to capture the variation of sound across different frequency. Therefore, this signal is very suitable for constructing the sound field around the watch.

It has a good detection effect for instantaneous gestures (e.g., wave, click) and continuous gestures (e.g., keeping wrist up, covering smartwatch with phone). It can provide rich audio features in the time and frequency domains. The frequency of the IMU is 200Hz, which records the acceleration and attitude angle of the smartwatch on three axes.
\begin{figure}[htbp]
\centering
\includegraphics[width=12cm]{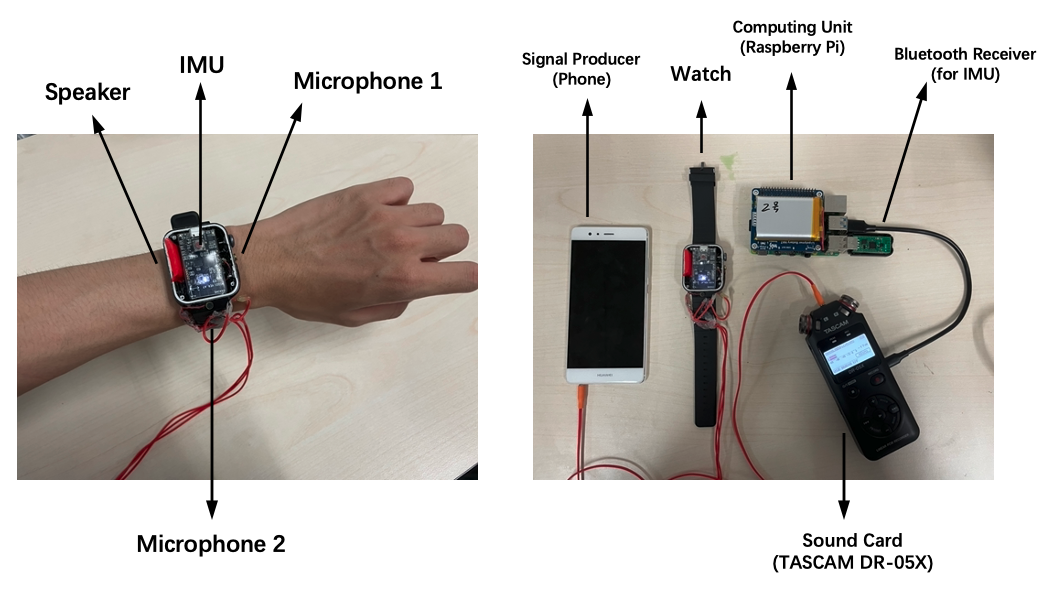}
\caption{The sensor distribution when the user wears the watch (left) and the complete \projectName{} hardware system (right)}
\label{fig_3_h_2}
\end{figure}

To realize the mobile computing capability, we connect the microphone to a TASCAM DR-05X recorder, make it act as a sound card to power the microphone, and import the signal output to the computing unit (Fig. \ref{fig_3_h_2}). The imu signal is sent to the computing unit through Bluetooth. The speaker is connected to the audio playback device to play the signal. The system uses MacBook Pro and Raspberry Pi 4B as computing units for prototype verification and mobile solution construction. There are differences in audio playback when using different computing units due to their hardware and performance. To maintain the stability of audio output, we use Huawei Mate 40 as the playback source. The device will continuously play audio with a 48khz sample rate through Hiby Music.

\subsection{Signal analysis}
\label{signal}
The principle of \projectName{} is similar to that of sonar: the speaker continuously emits a high-frequency audio signal, and the user's interactive behavior is sensed by analyzing the audio signals received by the microphone and the motion signals collected by the IMU. The user's gestures will affect the audio signal received by the microphone. Different gestures may generate audio reverberation or affect the loudness and energy of the signal in the time domain. The motion signal of the IMU has also been widely used in gesture recognition. We combine the two types of features to enhance the robustness and recognition range of the algorithm. We selected two representative gestures ( tilting the wrist up and clicking the microphone) from the pilot study for analysis.
\begin{figure}[htbp]
\centering
\includegraphics[width=14cm]{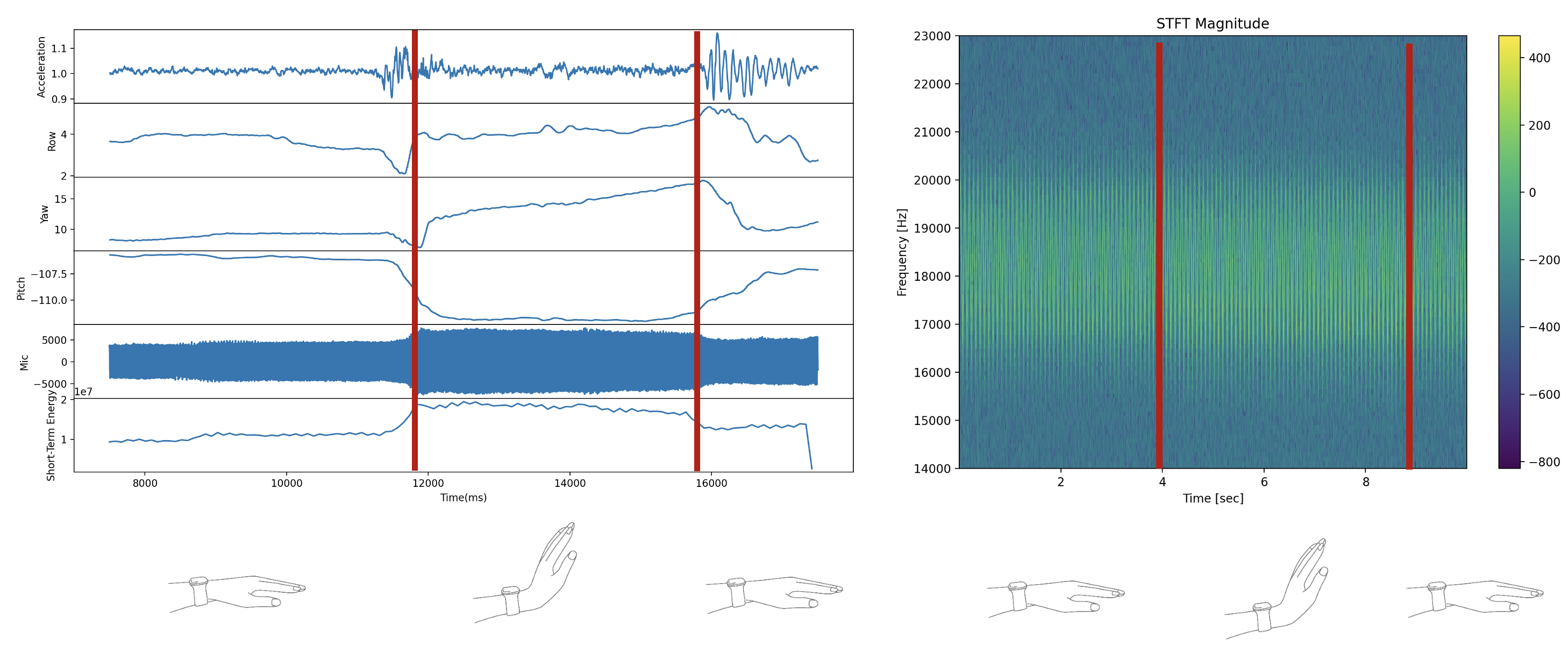}
\caption{The signal when the user performs \textit{Wrist Up}. The left image contains the overall acceleration of the three axes and Euler angle from the IMU, the amplitude and the short-term energy signal of the audio. The right image contains the STFT (2048 in each window) magnitude  of the audio signal. The red line in the figure divides the signal segments for the user to relax the wrist and tilt the wrist.}
\label{fig_3_p_1}
\end{figure}

\begin{figure}[htbp]
\centering
\includegraphics[width=12cm]{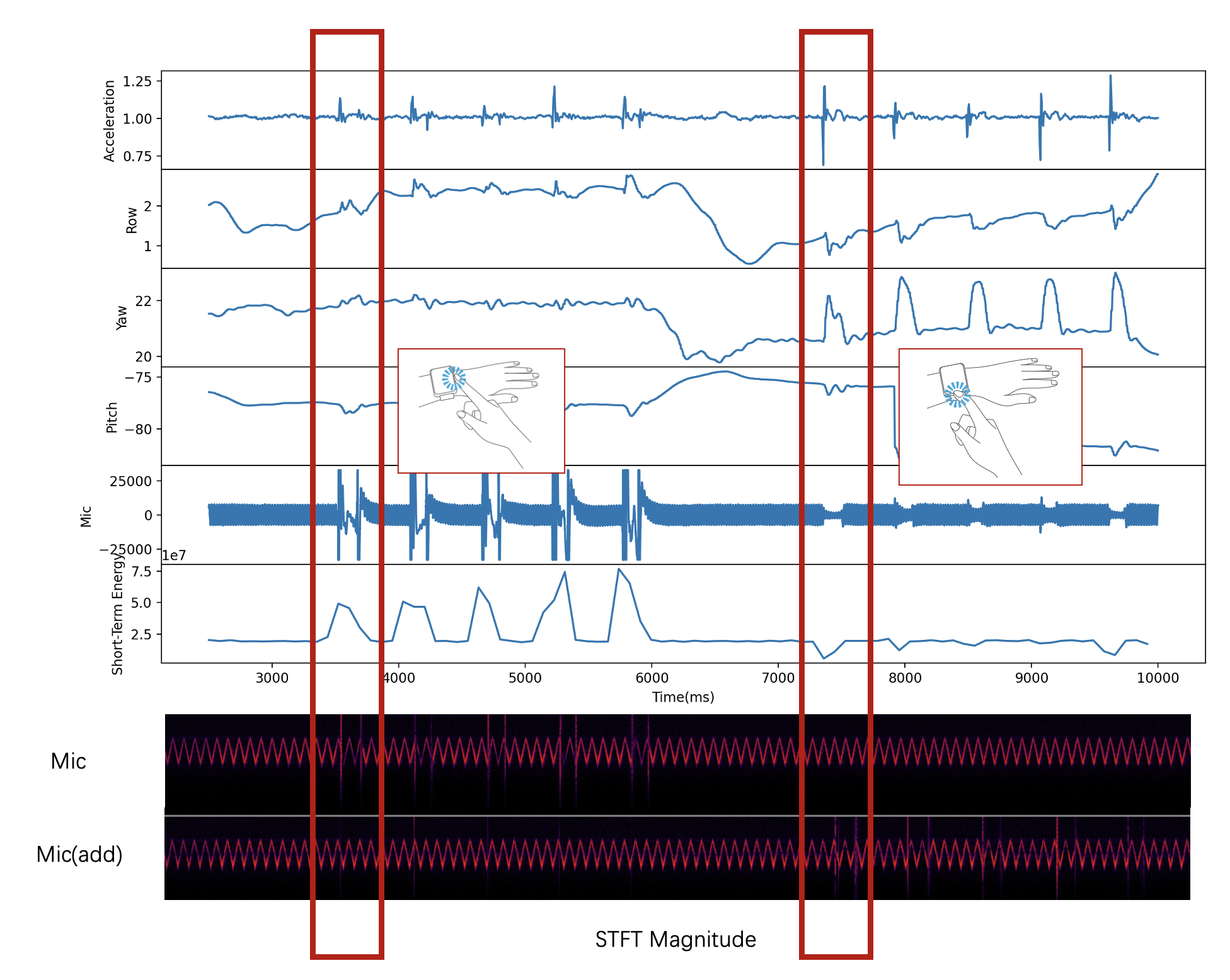}
\caption{The signal when the user taps two microphones for five times. The image contains the overall acceleration of the three and Euler angle from IMU, the amplitude, short-term energy, and the STFT magnitude of the two audio chanels. The two red boxes mark the one of the signal fragments that user performs a single gesture.}
\label{fig_3_p_2}
\end{figure}

Fig. \ref{fig_3_p_1} captures the motion information and audio changes when the user tilts his wrist up. The right pic shows the short-time Fourier transform (STFT, 2048 in each window) results. IMU's motion signal is stable when the user keeps the wrist relaxed and tilts it up. The imu only received an clear peak signal at the moment of tilting the wrist. For static gestures such as \textit{Wrist Up}, the user usually needs to hold this state for a while. During the process, the motion signal is always stable. When keeping the wrist up, the user's posture affects the transmission and reception of sound signals. As the figure shows, after tilting the wrist, the short-term energy of the sound signal significantly increases, which is also reflected in the energy of the frequency spectrum. The back of the hand reflects and absorbs the signal, which causes the change at different frequencies. The right figure shows that the frequency domain's energy distribution differs after the wrist is raised. Similar to raising the wrist, when the palm is occluded on the watch or the hand is in the pocket, the data received by the microphone will generate features with different characteristics. These features are related to the occlusion caused by different angles and materials.

Fig. \ref{fig_3_p_2} captures the motion and audio information when the user taps the right mic and the mic below the screen with his hand raised. Unlike static gestures, these dynamic gestures which performed in a certain period only have obvious signals when triggered. The acceleration has an obvious peak value when the user clicks the mic, and the attitude angle is also affected by the click and has a short-term fluctuation. The sound signal received by the microphone gets interfered with when the click occurs. When the user directly clicks on the right microphone, the sound reception is temporarily blocked, resulting in a gap-like loss in the continuous signal. When the user clicks the microphone below the screen, the signal received by the right mic also changes for a short time because the finger affects the conduction of the sound. Although this change is weaker than clicking directly on the right mic, it also has distinguishable features. At the same time, we found that dynamic gestures such as clicking always last for a short time and cause obvious peaks in acceleration. The signal can be captured using a specific size time window to complete the recognition of dynamic gestures.

\section{Interaction Design Space}
\label{design}
\begin{figure}[htbp]
\centering
\includegraphics[width=15cm]{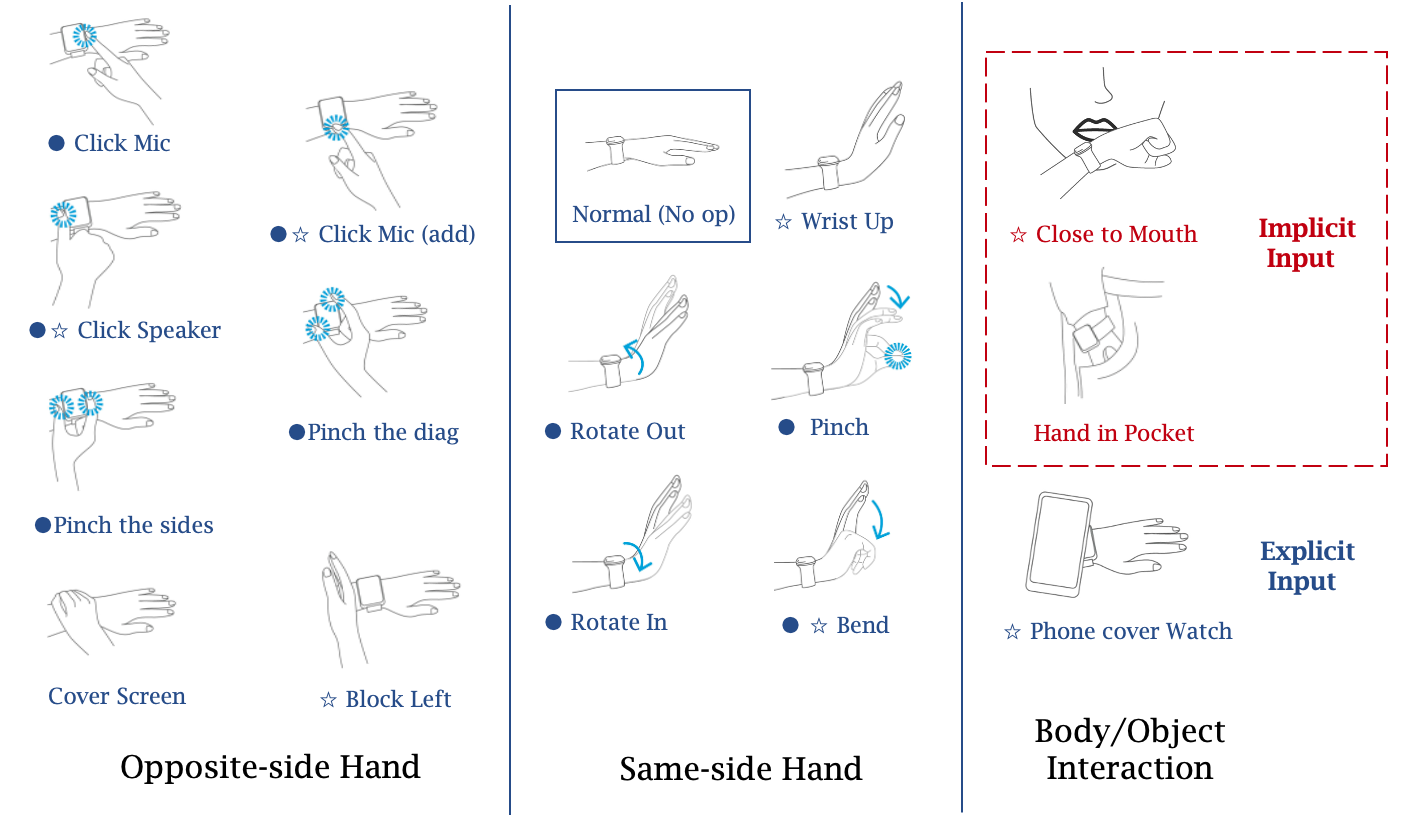}
\caption{15 gestures selected for \projectName{}. The gestures in the red box are the implicit input of the user, and the others are the user's explicit input. Dots mark dynamic gestures. Asterisks mark novel gestures.}
\label{fig_4_g}
\end{figure}

This section will introduce the gesture design space of \projectName{}. We analyzed the potential dimensions of smartwatch interactions and leveraged findings to construct a gesture set. We divide gestures into three subdivision types: Opposite-side Hand Gestures, Same-side Hand Gestures, and Body or Object Interaction. We selected 15 gestures as recognition targets and marked novel gestures not mentioned in previous studies in figure \ref{fig_4_g}.
The three gesture types correspond to different perceptual dimensions of the sensing field, and we will discuss their characteristics in detail below. The goal of this section is not to enumerate all possible gestures but to illustrate design possibilities and explore the potential capabilities of the acoustic sensing field. 

\subsection{ Opposite-side Hand Gesture}
Interaction with the opposite-side hand is the most commonly used method by users. The touchscreen interaction of smartwatches is also with the opposite-side hand. \projectName{} uses sound fields to extend the watch's sensing range to the around space, expanding the interactive range of the opposite-side hand. The design space includes the expanded touch, where users can touch beyond the screen, and the non-touch gesture users perform near the watch.

For touches, when the user does not operate, the signal received by the microphone can be divided into two parts: the solid-state sound inside the watch and the airborne sound outside the watch. When the user's finger touches different positions, the sound transmission path changes differently, producing prominent signal characteristics. An intuitive example is that most airborne sound is blocked when the user taps the microphone with a finger. Sound transmission path features play a crucial role in touch-related gesture recognition.
At the same time, when the user's hand touches the watch, a slight vibration will be generated. The IMU can capture these features to provide auxiliary information for identification.
We selected five different touch gestures, including taps on the sides of the watch and pinches on the edge of the watch, which users also recognized in previous studies.

Recognizing the characteristics of reflected sound is the basic principle of sonar.
For touch gestures, understanding transmission path characteristics is crucial, while when considering touchless gestures, different gestures will produce different reflected signals. Recognizing features from reflected sound
extends the sensation of the sound field from the watch into the space around it. 
The unique reflected sound signal generated at different positions facilitates the sensing of nearby gestures. 
Considering reflection features affected by sensing field size and hand position, we illustrate this with reflections near the three sensors (speaker and two microphones). 
This approach is demonstrated through three gestures: \textit{Wrist Up}, \textit{Block Left}, \textit{and Cover Screen}, discussed in detail in Section \ref{eval}.

\subsection{Same-side Hand Gesture}
The interaction of the same-side hand gives users quick commands when the opposite-side hand is occupied. Due to the field range constructed by \projectName{}, gestures of the same-side hand are difficult to affect the sound field directly. In the absence of other sensors, Serendipity\cite{serendipity} attempted to detect the motion signals generated by the same-side hand using IMU. But the average f1-score of 5 fine-motor is 87\%, which is difficult to meet actual needs. We considered enhancing the detection of the same-side hand using the sound field.

\textit{Pinch} and \textit{Rotate Wrist} are gestures that have been widely explored in many previous studies\cite{WristFlex, zhang2018fingerping, zhang2017fingersound}. One solution to detect same-side hand gestures is that the user can perform them based on the \textit{Wrist Up}. When the user tilts the wrist up, the arm will be more stable, and the motion signals of gestures such as pinch will be more obvious. At the same time, the \textit{Wrist Up} gesture ensures that same-hand gestures will not be easily triggered by mistake. We selected four gestures of \textit{Pinch}, \textit{Rotate-In}, \textit{Rotate-Out}, and \textit{Bend for testing}, which shows that \projectName{} can indirectly improve gesture recognition. We classify these four gestures as indirect gestures.

\subsection{Body or Object Interaction}
In addition to the signal influence brought by the user's hand, other body parts or objects can also influence the sound signal. This feature comes from two parts, one part is affected by factors such as the shape of the reflector and the reflection angle, and the other part is caused by the different sound reflection coefficients of different materials. For example, metals reflect sound more readily than woven fabrics.

Different objects have distinguishable absorption and reflection ability for sound, and their characteristics make the sensing field recognize the interaction with the hand and other objects or devices. We selected three typical gestures based on the usage scenarios of the watch: \textit{Close to Mouth}, which usually occurs when the user performs voice input;  \textit{Hand in Pocket}, often encountered when the user is wearing a smartwatch; \textit{Phone cover Watch}, an example of other devices interacting with the watch. In addition, using the watch close to objects, such as desktops and papers, also has potential research value. Here, we only selected three gestures to explore the potential capabilities of the sensing field built by \projectName{}.



The 15 direct gestures we selected include the static and dynamic gestures mentioned in section \ref{signal}(dynamic gesture marked with dots in the figure). We invite 15 users (8 male; the average age of 22.4; average smartwatch operation frequency is 6.6 times/day) to score these gestures based on a 5-point Likert scale from five aspects: Naturalness, Learning Cost, Fatigue, Social Acceptance, and Fun. All gestures have an average score of 3.5 or higher. Although some gestures are highly related to sensory ability, users still accept them.

\section{Implementation}
In this section, we will introduce the recognition algorithm of \projectName{} for all gestures proposed in interaction design. We collected data through a user study, designed a pipeline for data processing,  and finally obtained a trained model through machine learning methods.

\subsection{Data collection}

We designed a user study to collect data when users perform our selected gestures. The 15 gestures set contains six static gestures (\textit{Wrist Up}, \textit{Cover Screen}, \textit{Block Left}, \textit{Close to Mouth}, \textit{Hand in Pocket}, \textit{Phone cover Watch}), five dynamic gestures (\textit{Click Mic},\textit{ Click Mic(add)}, \textit{Click Speaker}, \textit{Pinch the sides}, \textit{Pinch the diag}) and four indirect gestures(\textit{Pinch}, \textit{Rotate In}, \textit{Rotate Out}, \textit{Bend}). Since users perform dynamic gestures differently, we also collected static data for all dynamic gestures. Take \textit{Click Mic} as an example. Some users will quickly click on the microphone and then take their hands away, while others may keep touching it for a while. We additionally collect data on the continuous contact of the user's finger with the microphone to ensure that both situations can be accurately identified. Due to the particularity of indirect gestures, we will collect their data separately. 

\begin{figure}[htbp]
\centering
\includegraphics[width=15cm]{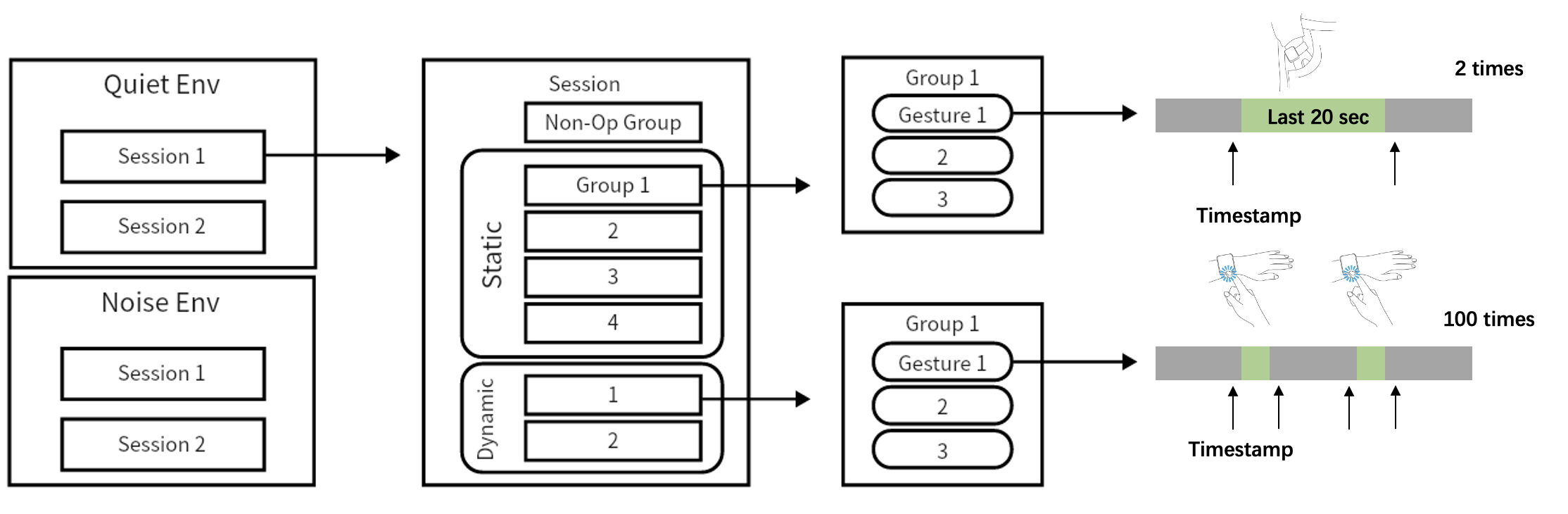}
\caption{The collection process for Part one. User will click Enter on the keyboard before and after each gesture to make timestamp.}
\label{fig_5_collec}
\end{figure}

The data collection study is divided into three parts. The first part is gesture data collection. We collected data for 11 direct gestures and non-operation data as false cases. For static data, we asked the user to hold the gesture for some time (20 seconds twice in one group) and recorded data. For dynamic data, we asked the user to perform the gesture many times (100 times in one group) and recorded the data. In order to reduce the burden on users, we collected data in groups, and each group contains three gestures. Due to the different acquisition methods of static and dynamic data, we did not mix the two gestures but randomized them internally. Each user needs to collect data from two sessions. Each session contains seven groups, one of which is the control group (to supplement the data when the user is not operating), four groups of static data, and two groups of dynamic data. All gestures are collected in random order.
The data collect method is shown in figure \ref{fig_5_collec}.
Each static gesture will be performed twice, each lasting 20 seconds. Each dynamic gesture needs to be performed 100 times($ 2\times 50 $ in one group) by the user. Before each gesture starts and ends, we ask the user to tap enter on the keyboard to record the timestamp. The devices for data collection were the same as the system built in the section\ref{hard}. To facilitate the data collection part, we used a MacBook Pro instead of a Raspberry Pi. Considering that \projectName{} is based on sound sense, we chose four different environments for our experiments. Taking the quiet laboratory as the primary data collection environment, we collected data in three different noise environments: outdoors, in shopping malls, and in restaurants. The collection equipment and collection process in different environments are consistent. 

The second part is subdivision gesture data collection. We collected data only on three gestures(\textit{Block Left}, \textit{Cover Screen}, and \textit{Wrist Up}) with different degrees. We collected data when the wrist is tilted up at three angles (horizontal, 45 degrees, and close to 90 degrees) and when the hand and the watch are at three distances (about 1cm, 2-5cm, and beyond 5cm). There are nine gestures to be collected in the second part. All of them are static gestures. Each user needs to collect data from two sessions. Each session contains three groups. The nine gestures are collected in random order. Unlike touch gestures, the signals of touchless gestures can vary greatly depending on how the user performs them. Considering the sensing principle of \projectName{}, when the tilt angle of the wrist is more significant, or the distance between the back of the hand and the watch is closer, the signal obtained by the sensor is more prominent. We subdivide the gesture according to the distance between the user's hand and the watch. The three levels of gestures correspond to the situation where the user has a clear intention, the user has a vague intention, and the user has no intention. We only use this part of the data to test the sensitivity of \projectName{} to touchless gestures. 

The third part is indirect same-side hand gesture data collection. We collected four indirect gestures and no-operation data with Wrist Up as false cases. Since all indirect gestures are dynamic, their collection method is consistent with the description in the first part. Each user needs to collect data from five sessions. Each session contains two groups. The first group is the control group, collecting the data of no-operation with \textit{Wrist Up} (20s $\times$ 3 times $\times$ for each gesture). The second group contains four gestures(100 times for each).

After each data collection group is completed, the user can rest. Before the next data collection group, the user must adjust or re-wear the watch. Because the two groups of gesture data are different, we did not force users to take off the watch. They can also adjust the position of the watch on the wrist or adjust the tightness of the strap. After the user completes one session, he is required to wear the watch to ensure the universality of the data. During the experiment, we did not require the user's posture. The user can complete the experiment in a standing or sitting position, and some users even walk around in a small area during data collection. The user's arm can be placed on the table or kept in the air during the experiment, which ensures that the data we collect has strong generalizability.

In the first part, we invited 12 people(6 males; average age of 23.2) for data collection. All of them need to complete two sessions in a quiet laboratory environment and also need to complete two sessions in a noisy environment. We collected data in three different noise environments, four persons per environment, and the order was random. In the second part, we invited six people(4 males, average age of 22.5) for data collection, all in a quiet laboratory environment. In the third part, we invited 16 people(9 male; average age of 22.1) for data collection, all in a quiet laboratory environment.
All the users wear the watch with their left hand.

We divided the data according to the timestamp recorded by the user in the study. We deleted 2s before and after each piece of static gesture data, considering the delay of the user typing on the keyboard and the delay of the system response. In the first part, we obtained a total of 1427 (98.08\%) pieces of valid data, 28 (1.92\%) pieces of incorrect data, 6024 (98.85\%) valid clicks, and 70 (1.15\%) invalid clicks. In the second part, we got 220 (96.07\%)pieces of valid data and 9 (3.93\%)pieces of error data. In the third part, we got 6216 (97.12\%)pieces of valid gesture and 184 (2.88\%)pieces of error data. The errors mainly come from two parts. One part is due to the user performing the wrong gesture or the time stamp recording problem. For example, the user taped the wrong microphone in the collection or forgot to make a timestamp while finishing the gesture. The other part comes from the device data recording problem. For example, the audio is missing a portion of data at the beginning or end of the record.

We carefully observed the user's performance during the study. Most users(11 out of 12) said they could not hear the high-frequency sounds from the watch, and one user said he could hear a slight chirp, but it did not affect his interaction. For the proposed gestures, the way different users performed is very similar. For example, in the absence of guidance, most users(10 out of 12) use the index finger to click the right mic and the mic on the band, and the thumb is used when clicking the speaker, which makes the data of different users more consistent. Some users initially chose to use their index finger to click on the speaker but replaced it with their thumb in the subsequent clicks. For some gestures, such as tilting the wrist up, some users choose to make a fist when the wrist is rotated upwards, while some users adopt an open palm. The diversity of these gestures may cause differences in signals, but the similarity of these gestures can still help us complete the classification. To ensure the naturalness of user interaction, we keep the gesture signals of different postures.

\subsection{Model Pipeline}
\begin{figure}[htbp]
\centering
\includegraphics[width=15cm]{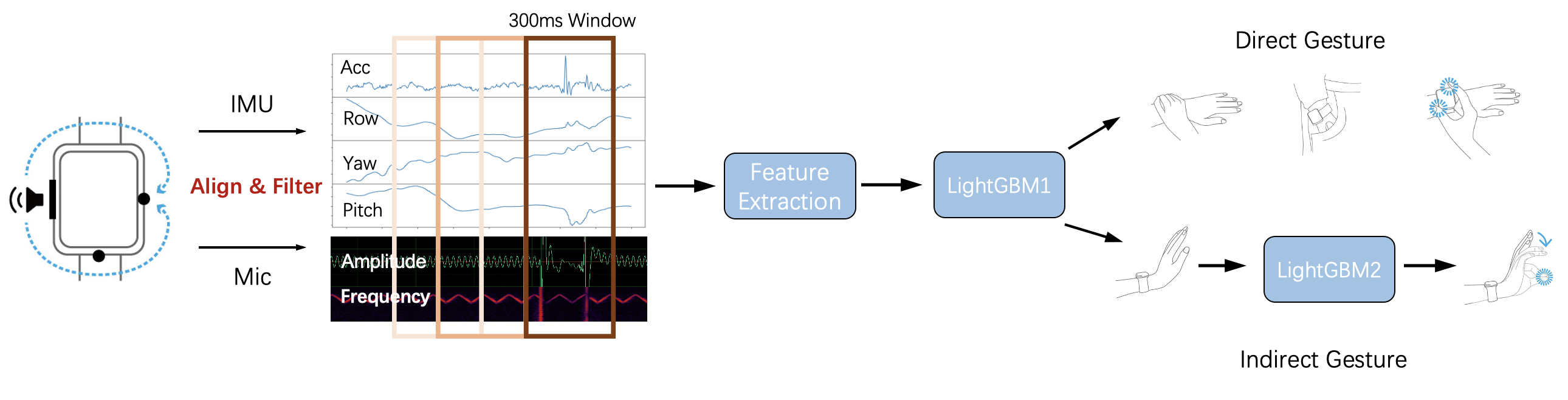}
\caption{The data pipeline of  Sonarwatch. The sensor data needs to be aligned and filtered, time window segmentation, feature extraction, and model recognition to obtain final gesture recognition results.}
\label{fig_5_a_p}
\end{figure}

We divided the \projectName{} recognition algorithm into three parts: data preprocessing, feature extraction, and model training. We use the method of moving the time window for real-time gesture recognition, so in the training data preprocessing part, we first use the time stamp of the data collected by the system to align the motion signal and the audio signal. Considering the difference in the sampling rate of the two signals, the sampling rate of the motion signal is only 200hz, which is much lower than the 48khz of the audio signal, so we require that the time difference between the audio and motion data in the time window does not exceed 5ms. We found that the time for a user to perform a single dynamic gesture operation does not exceed 300ms (99.5\%), and the holding time of static gestures is generally more than 300ms, so choosing a time window with a length of 300ms can meet the detection of two gestures at the same time. Each time window contains three complete chunks of binaural data and about 60 motion data. For the step size of the time window, if it is larger than 300ms, some gestures may not be detected, and if the step size is too small, it will increase the computational burden of the system. We finally chose 150ms for data detection. We segmented and annotated the collected static gesture data according to selected time windows and step lengths. For dynamic gestures, the place where the acceleration signal gets peaks is where the gesture occurs, so we calibrated the data around the peak of the acceleration and added two data points around the peak.

According to the analysis in section \ref{hard}, different gestures have different features on short-term audio energy and frequency distribution. After dividing the data into 300ms units, we perform feature extraction on the audio and motion signals separately. Since we used the original signal of 16.5k-20khz for the audio signal, we first filtered the received signal and only considered the high-frequency signal higher than 16.5khz in the calculation. For the three chunks (4096 sampling points in each chunk) contained in each data segment, we calculated their short-term energy, obtained the spectrum through a fast Fourier transform, and selected the spectrum part higher than 16.5khz to participate in the calculation. During data collection, we recorded the acceleration magnitude and the Euler angles in three directions for motion signals. We calculated the kurtosis, skewness, mean, variance, and maximum value of the four data types. These features can describe the overall stability and changes in motion data. The Doppler method is a commonly used audio-related gesture recognition method but is only effective when recognizing dynamic gestures. We did not involve the Doppler method to ensure that the model can recognize gestures simultaneously.

\begin{figure}[htbp]
\centering
\includegraphics[width=12cm]{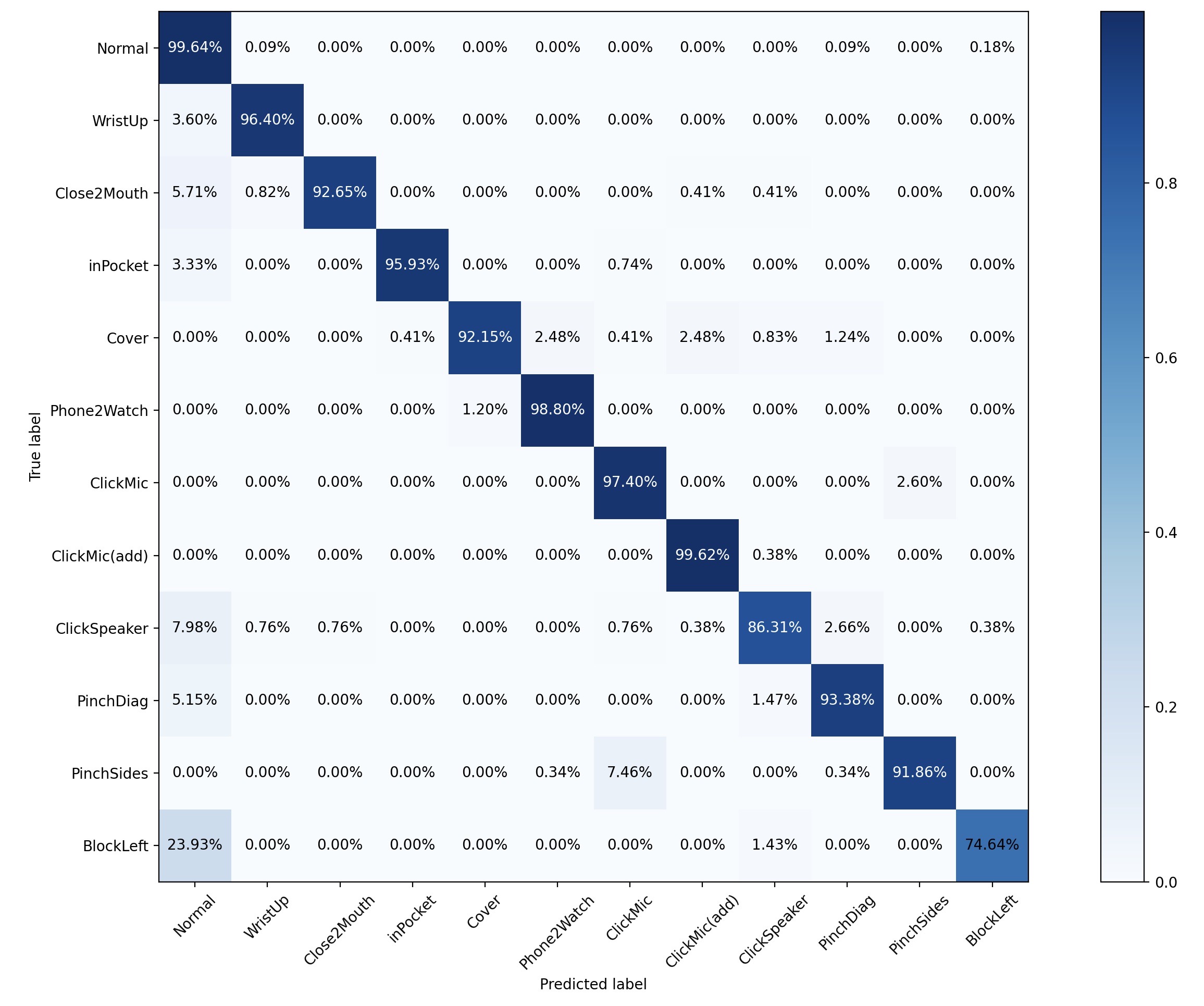}
\caption{Confusion matrix for recognition of 12 gestures (1 normal state, 11 user gestures)}
\label{fig_5_a_m}
\end{figure}
In selecting the final model, due to the limited computing power of the smartwatch itself, we chose the Light Gradient Boosting Machine \cite{ke2017lightgbm} as the classifier to reduce the energy consumption brought by the algorithm \cite{al2019comparison}.
LightGBM is based on the Decision Tree algorithm, which is fast and efficient, has low memory usage and high accuracy, and supports parallel and large-scale data processing. Smartwatches' computing and storage capabilities are generally lower than those of mobile devices such as mobile phones. We hope that \projectName{} does not rely on higher computing power or larger memory. LightGBM can meet our requirements for algorithm accuracy and computing power consumption at the same time.

\subsection{Overall Performance}
We mixed and randomly shuffled the data of all users. We divided the data into ten and divided the training and testing sets according to the ratio of 9:1. We obtained a ten-fold within-participant cross-validation accuracy of 93.7\% (with average f1 score 93.94\%). 

Since indirect gestures occur near the hand, they have little effect on the sound field of the watch at the wrist. The distinctive motion patterns accompanying these gestures facilitate their differentiation,
thereby enabling us to exclusively rely on motion data furnished by IMU for the categorization of indirect gestures. Given that the indirect gestures all based on \textit{Wrist Up}, we employ a two-stage recognition approach. We first detect the \textit{Wrist Up}, and then utilize the IMU data to refine gesture identification to enhance accuracy.
The no-operation and \textit{Wrist Up} data without gestures is segmented into negative examples for model training. We obtained a ten-fold cross-validation accuracy of 97.6\% for indirect gestures.

\begin{figure}[htbp]
\centering
\includegraphics[width=8cm]{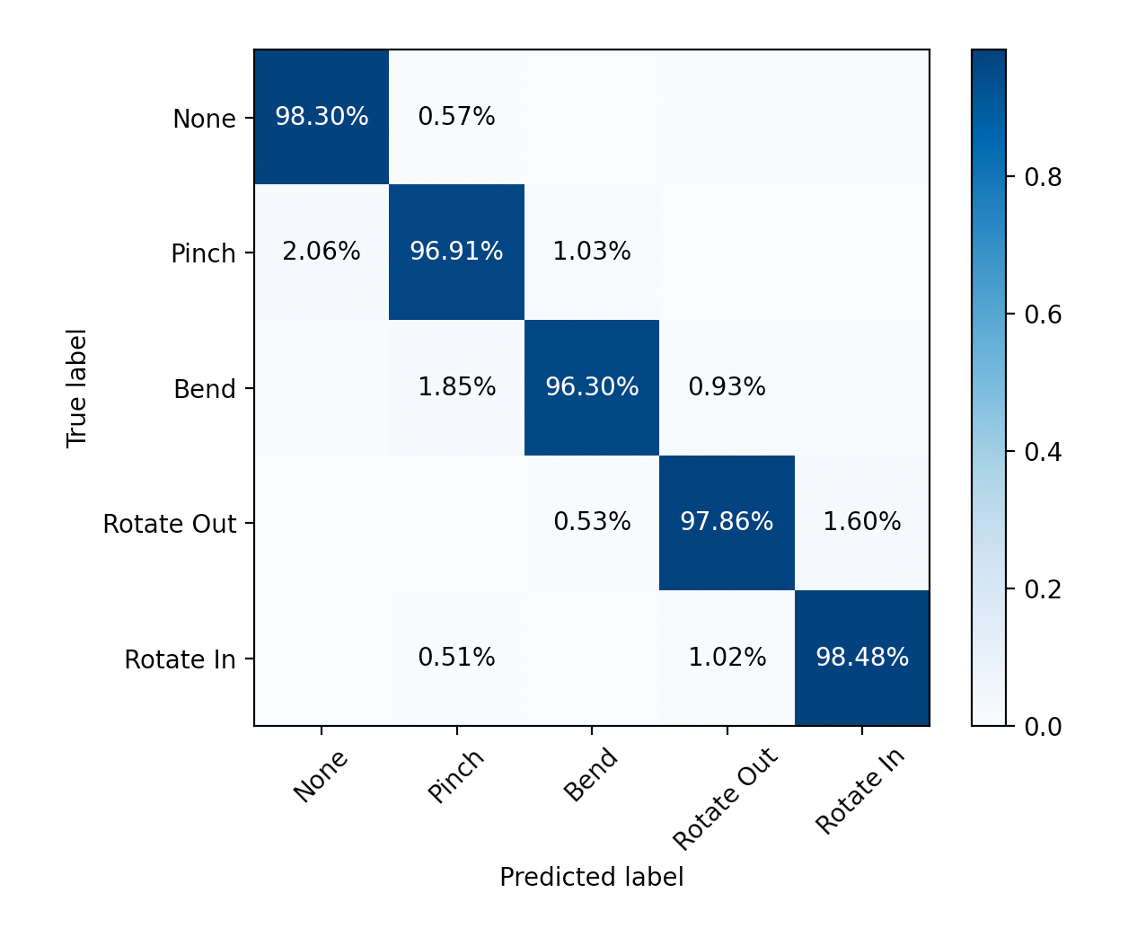}
\caption{Confusion matrix for recognition of indirect gestures}
\label{fig_5_a_m_d}
\end{figure}

The result shows that \projectName{} is rarely triggered by mistake, and a small amount of "false touch" happens on the two touchless gestures of \textit{Wrist Up} and \textit{Block Left}. From the confusion matrix, we can see that the accuracy of \projectName{} for opposite-side touch gestures is higher because the impact of touch gestures on the audio sensing path is noticeable, and gestures such as clicking on the microphone have irreplaceable characteristics in sound signals. The characteristics of touchless signals are more from reverberation and can easily get confused. Take the \textit{Wrist Up} and \textit{Block left} as examples, the characteristics of their sound signals are both reflected from the back of the hand. When the distance or angle between the user's hand and the watch changes, the degree of reflection will also vary. Due to the instability of the hardware itself, the signal received by the microphone may have obscure separable features. The worst recognition of Block also illustrates this, as many users may perform this gesture with the back of their hand far away from the watch, producing a poor signal with distinguishable features. Since different gestures occur in different watch directions, the sensors' distribution also affects the system's ability to perceive signals. Raising the Wrist has a noticeable reflection on the sound signal received by the microphone on the right side but has less impact on the strap microphone. In contrast, the Phone2Watch has a noticeable reflection on both microphones simultaneously, increasing the system recognition accuracy.
The accuracy of recognizing gestures using IMU in the \textit{Wrist Up} posture is significantly higher than the recognition accuracy in the standard posture (e.g., 83\% for \textit{Pinch} in Serendipity\cite{serendipity}). This disparity can be attributed to the stability of the user's \textit{Wrist Up} posture, proving the indirect use value of \projectName{}'s sensing field.

\section{Performance Evaluation}
\label{eval}
In the previous part, we found the factors that may affect the recognition ability of \projectName{}. In this part, we will evaluate the performance of \projectName{} in different situations.

\subsection{Different subdivision levels}
We selected as many gestures as possible in the interaction design section to explore the sensing range of the \projectName{} and obtained 93\% accuracy on the selected 11 gesture recognition. \projectName{} can sense rich interactive input. However, it's imperative to recognize that the gestures users commonly require for day-to-day activities tend to be straightforward. Obtaining a satisfactory level of accuracy on a small gesture set for daily usage holds greater practical significance. So we try to observe the performance of \projectName{} on a simplified set of gestures based on the same data from data collection.

We first test the accuracy of gesture recognition for different interaction methods, where the gesture for same-side hand interaction is based on the \textit{Wrist Up}. 
In particular, we observed that the recognition accuracy for body or object interactions was exceptionally high (99\%), which means that they have high practical possibilities. When users perform gestures with their hands, they raise their arms in a fixed posture, resulting in similar posture features. But the gestures associated with body or object interactions have more distinguishable features in the motion signal, making them easier to differentiate from other types of gestures.

\begin{table}[htbp]
\begin{tabular}{@{}cc@{}}
\toprule
Interaction Method   & Accuracy \\ \midrule
Opposite-side Hand Gesture       & 95.76\%   \\
Same-side Hand Gesture           & 97.6\%   \\
Body or Object Interaction & 99.1\%   \\ \bottomrule
\end{tabular}
\caption{Recognition Accuracy of Different Interaction Method }
\end{table}

In past research on watch interaction, touch, and non-touch interactions often require additional sensing support, among which touch interaction is more intuitive and easier to use. We mainly focus on \projectName{} in touch and non-touch performance under interaction.
Through the confusion matrix, we find that the recognition of \projectName{} for touch gestures is better than other gestures due to the more separable signal features of touch gestures. We proposed five touch gestures (\textit{Click Mic}, \textit{Click Mic (add)}, \textit{Click Speaker}, \textit{Pinch the diag}, \textit{Pinch the sides}) in the interaction design section. We tested the system performance only on touch gestures, and \projectName{} can achieve 98\% accuracy and 99\% for whether the user performs a touch gesture (Touch and touchless dichotomy). At the same time, we tested the other six touchless gestures with a recognition accuracy of 90.8\%.

\begin{table}[htbp]
\begin{tabular}{@{}cc@{}}
\toprule
Subdivision Level   & Accuracy \\ \midrule
Two Gesture Sets(No touch and Touch)       & 99.1\%   \\
Six Gestures(Normal and 5 touch gestures)           & 98.2\%   \\
Seven Gestures(Normal and 6 gestures) & 90.8\%   \\ \bottomrule
\end{tabular}
\caption{Recognition Accuracy of Different Subdivision Level Gestures}
\end{table}

Further, when considering the commonly used user interaction gestures,  \textit{Wrist Up} and \textit{Close2Mouth} were chosen as the targets for recognition. We excluded touch-related gestures as they are similar to those achieved by touching the screen or fixed buttons on the watch. \textit{Wrist Up} serves as the basis for same-side hand gesture recognition, while Close2Mouth is the pre-recognition gesture for the user's voice interaction. Recognizing these two gestures can greatly support same-side hand and voice interaction, making them highly valuable for recognition. The gesture recognition accuracy for three cases (\textit{Wrist Up}, \textit{Close2Mouth}, No-operation) reached 98.93\%. This high accuracy can effectively support users in completing smartwatch interactions using \projectName{}.

\subsection{Different sensor combinations}
In the implementation of the \projectName{}, we used the data of two microphones and IMU, but the current smartwatches on the market generally only have the microphone on the side of the watch, so we tested different sensor combinations for \projectName{}.

\begin{figure}[htbp]
\centering
\includegraphics[width=10cm]{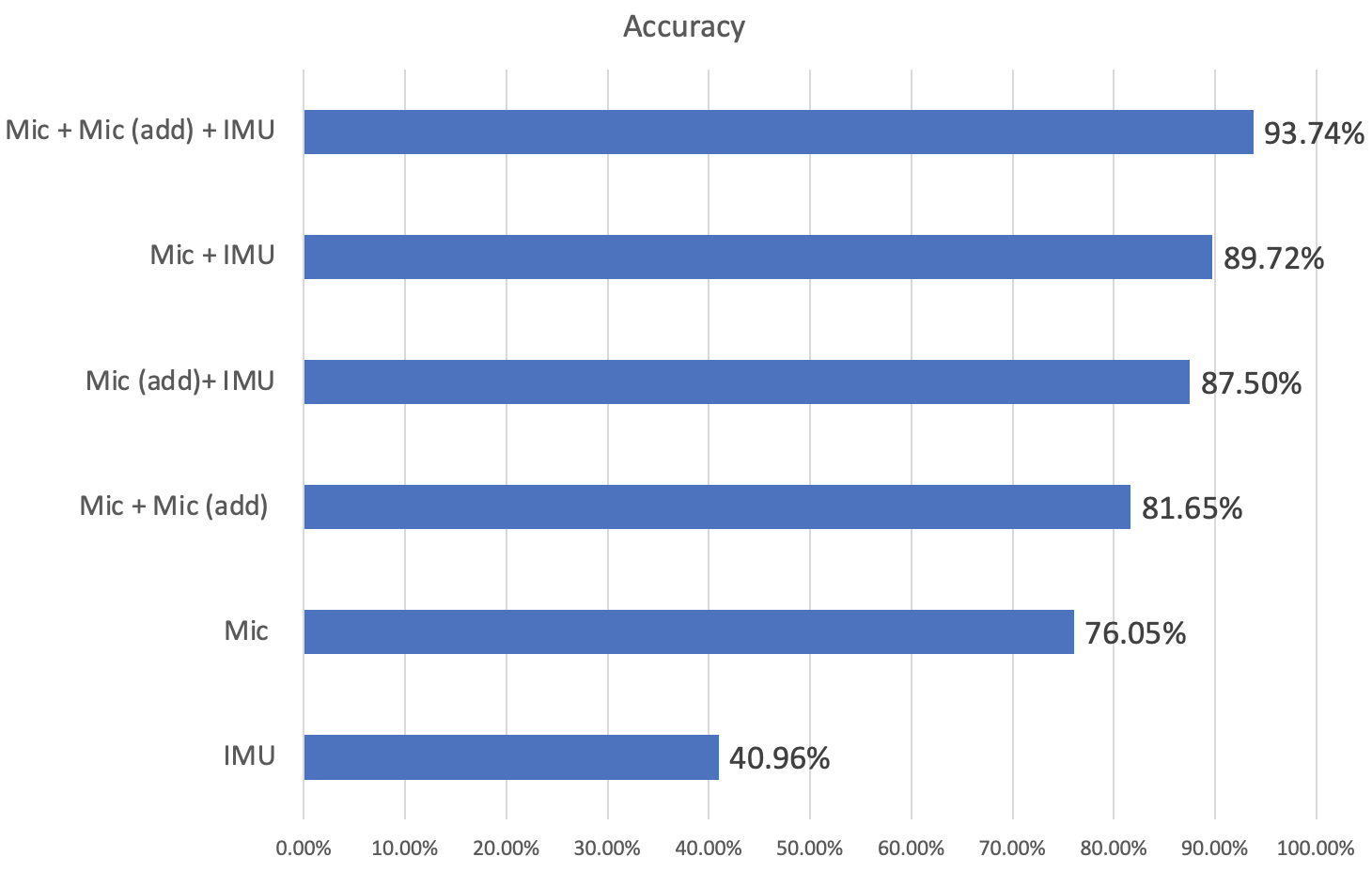}
\caption{The recognition accuracy of different sensor combinations}
\label{fig_6_e_c}
\end{figure}

The use of microphones in different positions shows the sensation advantages of different gestures. For example, for \textit{Wrist Up}, the audio signal collected by the right microphone has more obvious reverberation and energy changes, while for covering the screen, the microphone below the screen gets more obvious signal changes. It shows that a single microphone has limited perception of the sound field and is more sensitive to gestures in its vicinity. For the 11 gestures we proposed, the microphone performances of the two positions are close. Different microphone positions will affect the watch's ability to perceive different directions, and using multiple microphones simultaneously will increase the watch's sensation ability. We can choose a suitable position and combination of microphones for general situations according to the interaction needs.

We also tested the classification effect in the presence of IMU data. We ran RM-ANOVA tests ($p < 0.05$) on the accuracy of the ten-fold cross-validation, and the results show that the participation of IMU has a significant impact($F=33.851, p<0.01$) on the accuracy improvement. Without IMU, the recognition accuracy of \projectName{} dropped to 82\%. We analyzed the data and found that the confusion between \textit{Close to Mouth} and \textit{Wrist Up} increased when IMU was not used because the audio characteristics of the two gestures are similar. Most users(9 out of 12) have raised arms for around 40 degrees when performing the Close to Mouth, which has a more obvious feature in the signal captured by IMU. Using only audio signals increases false touches during the real-time test for gestures such as taps and pinch. We calculated the classification accuracy using only the IMU, and it dropped to 40\%, which is in line with our expectations. Because \projectName{} mainly relies on the change of the sound field to complete gesture recognition, the information provided by the IMU only plays an auxiliary role. However, there is a big difference in motion information in gestures such as \textit{Close to Mouth} and \textit{Hand in Pocket}, so only using IMU still has a recognition accuracy of 40\%.
In summary, the microphone provides a different level of sound field sensation capabilities, and the IMU provides part of the auxiliary information for gestures and judgment of the time point when dynamic gestures occur.

\subsection{Data sensitivity analysis}
\begin{figure}[htbp]
\centering
\includegraphics[width=10cm]{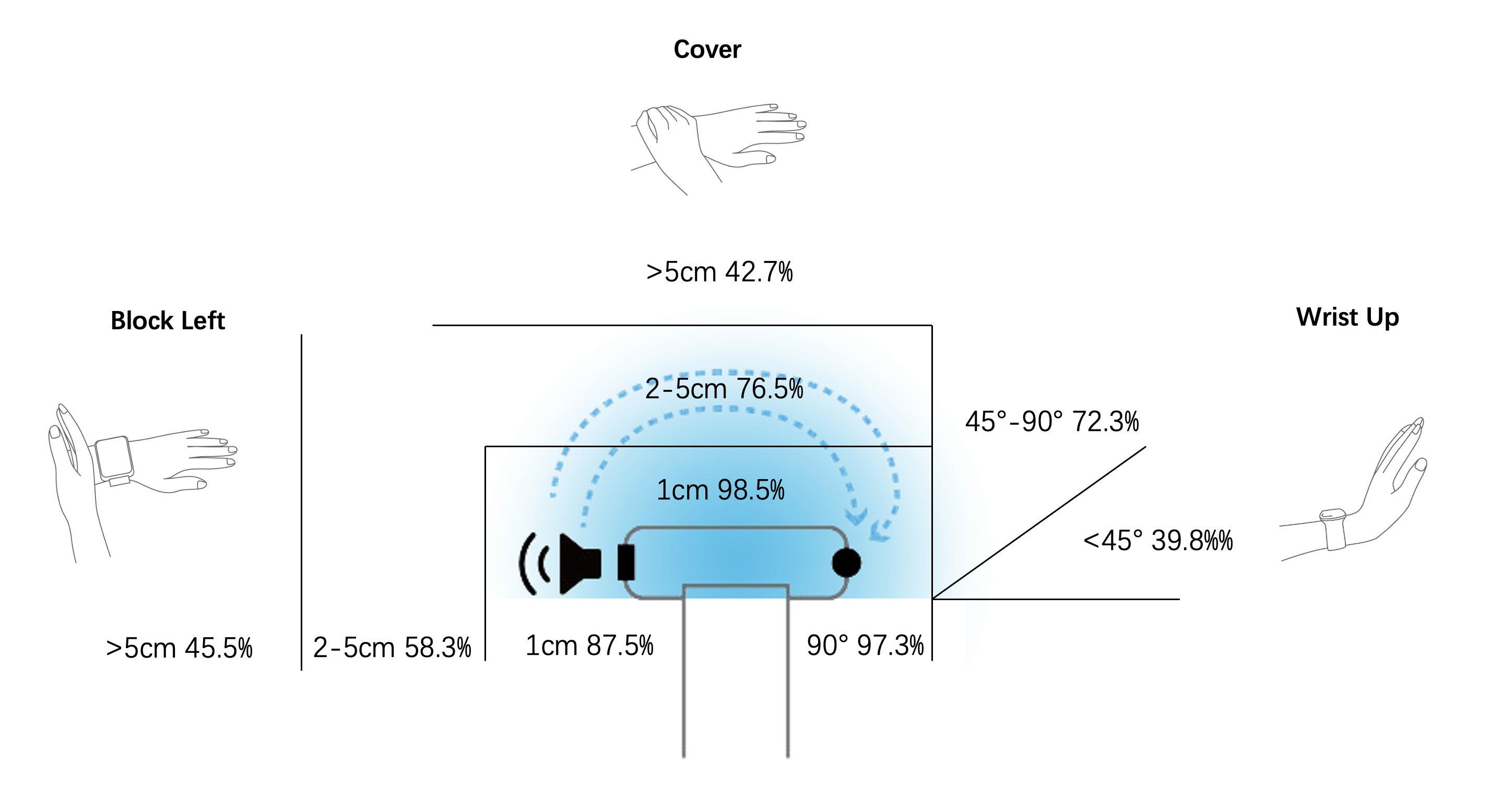}
\caption{Changes in the recognition accuracy of three gestures (Block left, Cover, Wrist Up)in different directions of the watch}
\label{fig_6_e_s}
\end{figure}

In the second part of the data collection, we collected three kinds of gestures with different accuracy ranges. These three gestures (\textit{Wrist Up}, \textit{Block left}, \textit{Cover}) represent the sensation accuracy of \projectName{} in three directions. When the sound signal in different directions is blocked, it will produce reverberation and energy characteristics. The closer the user's hand is to the watch, the more obvious the blocking effect of the sound. For example, for the \textit{Wrist Up} gesture, the sound signal is blocked by the back of the hand. When the user changes the angle between the hand and the watch, the audio signal will get different changes. A notable feature is that the audio energy received by the right microphone is higher when the hand is closer to the watch. The case where the user clicks the speaker can be regarded as a particular case where the hand contacts the watch, and the user does not perform a gesture can be regarded as a particular case where the hand is infinitely far away from the watch. We discuss the situation at different distances to understand the sensation accuracy of the \projectName{} in three directions, and the results are shown in figure \ref{fig_6_e_s}. 

As the distance between the hand and the watch increases, the \projectName{}'s recognition accuracy for gestures gradually decreases. \projectName{} is very sensitive to user gestures. The sound has obvious reverberation characteristics when the user's hand is close to the watch (about 1cm). When the distance between the hand and the watch is medium (2-5cm), the recognition accuracy of Cover and \textit{Wrist Up} dropped to around 75\%. At the same time, \textit{Block left} gesture's recognition accuracy was lower than 60\%. Since sound energy emitted by the watch is limited, the farther the hand is from the watch, the smaller the reflected sound energy is and cannot be sensed by the microphone. The gestures performed directly near the microphone have more separable reflection characteristics than those performed near the speaker. When the hand is 5cm away from the watch, the recognition accuracy drops to 50\%, and the system can hardly recognize the gestures at this level. 

\subsection{Noise Environment Analysis}
Since the \projectName{} is a sound-based technique, we tested its performance in different noise environments. We selected three typical scenarios with different noise levels ( outdoor, shopping mall, restaurant). The noise of the whole frequency band in these three environments is 34.45dB, 51.41dB, and 65.29dB (17.44dB in the laboratory environment), which can cover most of the usage scenarios of smartwatches. We collected data from four users in each environment during the data collection.
The signal's frequency for constructing the sound field is higher than 16.5khz to enhance the anti-interference of \projectName{}. When the noise signal of 16.5kHz-20kHz is not included in the ambient noise, other sound signals do not interfere with the \projectName{}. The test results show that the recognition accuracy of the data in the three environments is 93.15\%, 92.70\%, and 92.62\%. This is comparable to the recognition accuracy we obtained in the laboratory setting. RM-ANOVA test shows that there is no significant difference in recognition accuracy in different noise environments($F=0.849, p=0.516$). \projectName{} received little interference from noise signals, which also met our expectations.
\begin{table}[htbp]
\begin{tabular}{@{}clc@{}}
\toprule
Environment   & Noise (dB)  & Accuracy \\ \midrule
Lab           & 17.44       & 93.7\%   \\
Outdoor       & 34.45       & 93.15\%   \\
Shopping Mall & 51.41       & 92.7\%   \\
Restaurant    & 65.29       & 92.62\%  \\ \bottomrule
\end{tabular}
\caption{Recognition accuracy in different noise environments}
\end{table}


\subsection{Real-world Performance and Negative Example Analysis}
To evaluate the performance of \projectName{}, we used the training model obtained by the algorithm to build a real-time system for testing. We invited eight users (four male; average age of 20.5; none of them is involved in data collection) to wear watches for interaction. The user will complete the gestures the experimenter gave in a sitting and standing pose. We have no restrictions on the placement of the user's hands. Users can hang their hands or place them on the table when not performing operations. Each gesture was repeated ten times and in random order in each pose. We played 30dB of white noise (normal speech and ambient sounds) during the experiment to simulate a more realistic environment. We recorded the recognition results of the system during the study and the false touches that occurred in the interaction. 
We finally got a recognition accuracy of 90.5\% for all executed gestures. From observation, we found that for the recognition of static gestures, the system may give wrong results in the first three to five time windows. Since the user will continue to hold the gesture for a while, this error can be automatically corrected by the system in the subsequent recognition process.

While caring about the overall accuracy of the \projectName{}, we also care about the system performance for incorrect responses and negative cases. During the experiment, we recorded a total of 103 times of false touches when the user raised his hand, including 81 touchless gestures. We did not count the false touches when the user did not raise his hand because the user will not interact with the watch now. Most of the false touches were recognized as static gestures in a short period. To reduce false touches, we added a layer of judgment to the recognition algorithm. The recognition is successful only when three consecutive time windows are recognized as the same gesture, which can filter 40.8\% of false touches(42 out of 103). We have found two reasons for negative cases and false touches that the current algorithm cannot eliminate. One is from the hardware system. Due to the problem of sensor fixing and wire welding, when the user's arm moves, it may cause instability of the speaker and microphone to send and receive signals, resulting in false touches.
A typical example is that the audio signal received by the system fluctuates when the leads are squeezed due to the long leads of speakers and microphones, but this is rare. Another part comes from the algorithm itself. As we mentioned in the data collection section, how users perform gestures creates differences that affect the characteristics of the signal. At the same time, unexpected user behaviors may bring signal characteristics similar to existing gestures. For example, when some users raise their wrists, the back of their hands touches the microphone, which generates a signal similar to clicking on the microphone, leading to recognition errors.


\subsection{Overall algorithm energy consumption}

We tested the energy consumption of the watch while doing calculations by sending data and models to the Mi watch. We compared the power consumption of different apps. Since \projectName{} is based on sound sense, we compared its energy consumption with music playback applications. Here we choose QQMusic on Mi Watch as a comparison application. We ran the heart rate detection, QQ music, and \projectName{} detection separately in the same environment to observe the power consumption of the watch after continuously running for 40mins. We get the running power through the parameters given by the watch. The test results are shown in fig. \ref{fig_g_e_c} below.
Given the challenge of directly measuring the power of a running application, we resort to the power usage of the application during its operational phase as a indicator of its power consumption.
\begin{figure}[htbp]
\centering
\includegraphics[width=10cm]{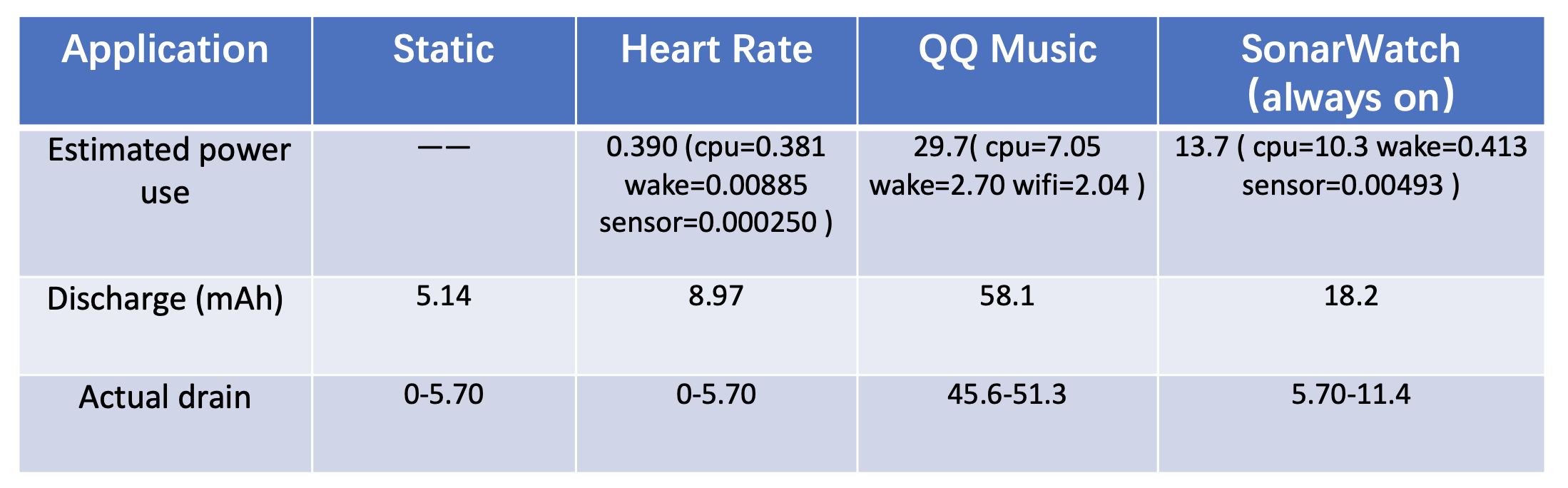}
\caption{The power consumption test results of the watch running different applications within 40mins. Estimated power use: The power consumption of the application-specific module is obtained by the system interface. Discharge: Calculated power consumption within 40 minutes. Actual Drain: Actual power consumption percentage within 40mins}
\label{fig_g_e_c}
\end{figure}

When the recognition algorithm works continuously in the background, its energy consumption is about a quarter of that of QQ Music. According to the power consumption results, the \projectName{} can be used continuously for 6 hours by the user. Considering that the user does not operate the watch all the time, assuming that the user operates the watch for one hour each day, the power consumption of using \projectName{} for interaction is about the same as the power consumption of continuous 24h heart rate detection.

\section{Discussion}
\label{dis}
In this section, we discussed the application opportunities and scenarios of \projectName{} and analyzed the limitations and future work.

\subsection{Interaction Design Extension}
We consider gestures highly related to \projectName{} as recognition targets in selecting the gesture set. 
In section \ref{design}, we discussed \projectName{}'s recognition enhancements for pinch and rotate wrist gestures. The user can tilt the wrist up and then perform pinch and wrist rotation while keeping the wrist raised, which does not bring more fatigue and can reduce false touches for the gestures. Raising the wrist can clarify the user's input intent and make gesture execution more stable. Furthermore, the static gestures detected by \projectName{} can be extended by more gestures. For example, after the phone is close to the watch, the user can use the phone screen to operate the watch or wake up the shortcut application. Users can use the middle finger to tap the microphone and the index finger to provide additional semantic touch input. The smartwatch can give feedback when the user's mouth is close to the microphone to remind the user that voice input is available, etc.

\subsection{Improvement of System and Signals}
The hardware system we built referred to some existing smartwatch sensing solutions on the market and tested the effect of adding a microphone. The evaluation results showed that the position and number of microphones have apparent differences in the ability of gesture recognition. In addition to the sensor distribution we used, existing smartwatches also have different distributed speakers and microphones. In addition, in this paper, we require the user to wear the watch with the screen facing up on the left hand, but the user may wear the watch with the right hand or with the screen facing down in actual use. 
The industrial design of the watch and the different wearing methods of users bring multiple sensor distributions, and these distributions bring possibilities for more possible sensing abilities and interaction designs. When the user is operating the watch, the clothing generally does not produce apparent occlusion, so we did not consider the case of the sleeve covering the watch. In some cases, because the user's clothes are too thick or the sleeves are too long, there may be interference with the sound field, reducing the accuracy of gestures \cite{nute197363}.
Besides, we chose the chirp signal as the ultrasonic signal to construct the sound field. Other works have different signal implementation schemes, such as using OFDM signals to enhance the spatiotemporal characteristics. The specific signal that can achieve a better recognition effect needs to be tested. We only give one possibility in this paper. We explored the interactive extension provided by sound field sensation for smartwatches, so the use of signals may be one of the future research directions.

\subsection{Application Opportunity}
We discussed the application possibilities of \projectName{} in other scenarios in addition to enhancing the current interaction experience of the smartwatch. We chose accessibility services and cross-device interaction for analysis.
The touch problem of the small screen significantly impacts the accessibility service based on Talkback on the watch, which makes it really difficult for the visually impaired to use the smartwatch. \projectName{} can provide simple gestures for visually impaired people to support the input of shortcut commands during daily smartwatch experiences and improve the efficiency of smartwatch interaction for visually impaired people.
In addition to being a watch interaction technology, \projectName{} can also complete the control between the watch and other devices. The gestures recognized by \projectName{} can be used for control commands of other devices or scenes, such as using \projectName{} to control mobile phones or as auxiliary input means in XR scenes. 

\section{Limitation and Future Work}
Finally, we review our work and discuss limitations and directions for future research. We cannot modify the AutoGain settings of the Android system. Modifying this permission requires high programming costs, so we use our hardware system for experiments. The signals received by our system may be different from those collected by existing watches.
We mainly explored the user's explicit input, but we did not deeply explore activity detection. We hope to study the sensation ability of the \projectName{} for the user's implicit input in future work. We only explored discrete gestures in the input stage. Considering the performance of continuous control technologies such as Tilt in smartwatch research, the continuous input combining Tilt and \projectName{} is also of great research value.


\section{Conclusion}
\projectName{} is a field-sensing smartwatch gesture interaction technology that can recognize various user inputs without additional sensors. The evaluation shows that \projectName{} can achieve 93.7\% 12-category recognition accuracy and 97.6\% for same-side hand gestures. \projectName{} has a wide recognition range, high deployment flexibility, sensitive sensing, and acceptable energy consumption. \projectName{} can be used for quick input of smartwatches, improving the efficiency of existing gesture input, optimizing accessibility services, and supporting cross-device interaction and control. We predict that \projectName{} will be welcomed and loved by smartwatch users and has great potential for interaction.


\bibliographystyle{ACM-Reference-Format}
\bibliography{sample-base}


\end{document}